\documentclass[referee]{mn2e}
\usepackage{graphics}
\title[Synthetic UIB spectra]{Synthetic spectra of the UIBs (2 to 40 $\mu$m)}
\author[R. Papoular]{R. Papoular$^{1}$\thanks{E-mail:
papoular@wanadoo.fr}\\
$^{1}$Service d'Astrophysique and Service de Chimie Moleculaire,\\
CEA Saclay, 91191 Gif-s-Yvette, France}
\begin{document}

\date{Accepted . Received ; in original form }

\pagerange{\pageref{firstpage}--\pageref{lastpage}} \pubyear{2002}

   \maketitle
\label{firstpage}

\begin{abstract}
Computational chemistry is used here to build a set of carbonaceous structures whose combined spectra approximately mimic typical UIB (Unidentified Infrared Band) spectra. A large number of relatively small hydrocarbon structures, containing traces of heteroatoms (oxygen, nitrogen and sulfur) were considered, including aliphatic chains, compact and concatenated hexagonal and pentagonal rings. Their ir (infrared) spectra were computed using standard chemistry software. Those which exhibited at least a few lines falling within one of the UIBs, and no significantly strong line outside the observed bands, were retained: in all 35 structures, grouped in 8 families and totalling about 6000 vibrational modes together. Each family exhibits a characteristically different spectrum. Guided by the IRS spectra of the Spitzer satellite, each of the 8 families was given a weight, which was tailored so that the concatenation of all 35 weighted spectra resembled UIB spectra. A typical chemical composition is found to be C:H:O:N:S=1:1.15:0.064:0.0026:0.013. The present procedure allows each structural family to be preferentially assigned to an observed UIB, which helps figuring out the structure of interstellar dust. The essential role of heteroatoms is apparent.

\end{abstract}





\section{Introduction}

To this day, the composition and structure of the carrier(s) of the Unidentified Interstellar Bands (UIBs) remain a subject of investigation and discussion. Obviously, the ultimate solution to this problem should be the collection \emph{in situ} of dust samples in various astronomical environments, followed by their analysis in the laboratory. A great stride was made in this direction with the return, in 2006, of samples from Comet 81P/Wild2 (Sandford et al. \cite{san06}). Preliminary results were discussed in several contributions to the IAU Symposium 251 in Hong Kong (Kwok and Sandford \cite{kwo08}). Work on these samples requires considerable technical means and sophisticated instruments; it is still actively going on.

While these and other similar efforts are being pursued, it is of interest to approach the problem from other sides and with different means. Several attempts have been, and are being, made to this end. The most popular is based on the use of measured or computed polycyclic aromatic hydrocarbons (PAHs) (see Bauschlicher et al. \cite{bau09}, Mattioda et al. \cite{mat09}). One of the latest examples of such a line of attack is given by Boersma et al. \cite{boe10}. It consists in drawing spectra from a growing data bank of PAHs of different shapes and sizes. In general, these are compact or irregular, neutrals, anions or cations, mostly of large size.

Another type of investigation, opposite in a way, was initiated by Papoular et al. \cite{pap89}. It is based on the analogy between some of the UIB spectra with some of the spectra of "standard" coals (see Charcosset \cite{cha}) and kerogens (see Durand \cite {dur}), collected in the rich data banks established over the years in France and the USA, by institutions dedicated to coal and oil mining, like the Groupement Francais d'Etude des Carbones (GFEC) and the Institut Francais du Petrole (IFP). This analogy was previously noticed by Kerridge \cite{ker87}. Analogy between kerogen and meteoritic organics was also noticed by planetologists (see e.g. Khare et al. \cite{kha}) and petrologists (see Cataldo et al. \cite{cat04}). An interesting feature of this method is that a tight parallel can, in some cases, be drawn between series of terrestrial and astronomical spectra (see Papoular \cite{pap01} for the CH stretching spectral region). Since laboratory and field studies allowed the changes in the terrestrial spectra to be ascribed to definite physical and chemical modifications, some of these conclusions can, by analogy, be extended to the IS band carriers, thus drawing a link between otherwise unconnected spectra and simultaneously clarifying their structure and composition.

This approach also highlighted the key role of heteroatoms in generating a number of features: oxygen, nitrogen, for instance, which, together with hydogen and carbon, form the CHON family, celebrated earlier on by planetologists.

While this approach helps in the fingerprints spectral region, i.e. below about 10 $\mu$m, where bands are clearly distinct, and assigned to definite small groups of a few atoms each, it is less powerful beyond that. Besides, even in the laboratory, the assignment of particular structures to longer wavelength bands has proved challenging (see Speight \cite{spe}). In the present work, we take advantage of the prodigious progress made, in the last two decades, both in computer technology and fundamental computational chemistry, to complement laboratory experiments with numerical determination of the ir (infrared) spectrum of several structures relevant to the kerogen  and coal models (see Carlson \cite{car92}). Calculations which previously required complex dedicated programming and long cpu times on powerful machines can now be performed in a few hours with state-of-the-art desktops and general purpose software (e.g. Hyperchem and Momec; see, for instance, Hypercube \cite{hyp}).

Moreover, such software delivers, for each mode, the ir intensity and graphic illustration of the movement of each atom in the structure, together with the frequency of vibration. This is of great help in selecting chemical elements and molecular structures of interest, and later estimating their relative abundances in IS dust.

Based on the availability of such new tools, the present work is an attempt to determine a set of simple chemical structures whose combination can reproduce the observed interstellar spectra and explain the variations of the relative band intensities from sight line  to sight line. The choices are constrained by the known relative abundances of elements and the (less well known) abundances of free-flying molecules. Another, self-imposed, constraint is to limit the elementary structures to the minimum in number and complexity. The aim is to determine the quantitative contribution of each  structure to the various bands, and, if possible, the type of molecular vibration which is activated.

Section 2 describes a number of types of elementary structures, or molecules, which contribute to one or more of the main IS bands. In this respect, the recent availability of observational data with increased spectral span, resolution and sensitivity from various celestial environments is a precious asset (Spitzer SINGS; see Kennicutt\cite{ken03}). For any proposed dust component must not only contribute to the observed spectrum, but also be proved not to contribute lines or bands that are not observed throughout the ir spectrum.

Individual small structures at low temperatures produce only narrow lines and cannot provide the observed band widths and continuum by themselves. However, small changes in the anchoring points of peripheral groups of atoms slightly shift the original lines, thus contributing to filling the band width without altering the topological character of the structure. The latter defines the type, or class, of the elementary structure. 

The concatenated spectra of 5 such classes reveal several clusters of ir lines, listed in Sec. 3, which are characterized by their vibrational mode, i.e. the atomic elements or functional groups involved in the motion, or the type of coherent motion of the atoms. Each cluster is found to coincide reasonably well with one or the other of the tabulated UIBs.

But the density of lines in each cluster is still insufficient for the synthesized spectrum to approach observed spectra. Section 4 therefore discusses the effects of linking together two or more of the selected elementary structures to give so-called composite structures. This, too, ``densifies" the line clusters, and so helps building up the spectral bands. It also gives rise to new ``global", or bulk, modes, farther and farther in the red, which contribute to the underlying continuum. It turns out that at least a few thousand atoms are required for the synthesized spectrum to compare with observed spectra. Other considerations (scattering cross-section, resilience, etc.) also point to much larger dust grain sizes ($\sim0.1 \mu$m). Based on the above, Sec. 5 displays the concatenated spectrum of a large family of elementary and composite structures, in adequate relative numbers so as to best fit a typical UIB spectrum. The corresponding dust composition and structures are given in Sec. 6.
 Finally, excitation and subsequent ir emission are discussed in Sec. 7.

\section{The elementary structures and their spectra}

This section describes the elementary structures that were used to build up a compendium whose ir spectrum looks like a generic UIB spectrum. They were selected for their atomic composition to be compatible with cosmic abundances (mainly hydrocarbons), for their simplicity and for their spectrum to display one or more lines within one or more UIB, but none outside. Clues were also found in the sketches of ``young" kerogen structures (see Behar and Vandenbroucke \cite{beh}, and coal structures (see Speight \cite{spe}). The latter include not only small PAH's, but also concatenated 6- and 5- membered rings. Our's is only an example of relevant selection, and does not, in any way, imply  that it is the best, nor that it is exhaustive.

The chemical computations were performed by means of HyperChem 7, a software package commercialized by Hypercube, Inc. The semi-empirical, PM3/RHF computation method was preferred because it was specifically optimized for hydrocarbon structures and gives sufficiently accurate ir freqencies (better than about 5 $\%$) within reasonable computation times. Molecular Mechanics methods are faster but inaccurate, while \emph{ab initio} methods cannot cope with the size of relevant structures (see discussion in Papoular \cite{pap01}). 

Each structure was graphically designed on a PC screen, then optimized for minimum potential energy. At this stage, the software can start computing the vibrational modes and their ir intensities. 
Figure 1 shows the following typical structures.

\begin{figure}
\resizebox{\hsize}{!}{\includegraphics{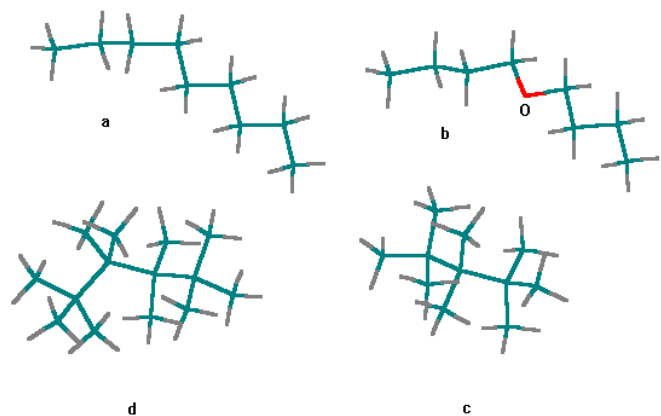}}
\caption[]{Chain structures. In agreement with chemical conventions, here and below, the bonds are represented by sticks, the nodes are occupied by C, O, N or S atoms, the free ends are occupied by H atoms. Color code for this and the following figures: green:C; gray: H; red: O; blue: N; yellow: S. a) Chain CH$_{2}$, 29 at.; b) Oxygen bridge, 27 at.: the chain is interrupted by an O atom; c,d) Chain CH${3}$a,b with 35 and 44 at.: here, there are no CH$_{2}$, and all terminations are methyls. }
\end{figure}

\subsection{Chains}

a) {\bf Chain CH$_{2}$} (29 atoms): a chain of a few CH$_{2}$ (methylene) groups linked by single covalent bonds. All C sites are sp$^{3}$, which makes this an aliphatic chain. Apart from the end methyls (CH$_{3}$), this type is expected to link together the more compact structures (see below) to form a dust grain in space. Its spectrum (in Fig. 5) is sparse and weak; however, its strongest features may contribute significantly near 13 $\mu$m. They are generated by  vibrational modes in which the methylenes of the linear part of the carbon skeleton are rocking perpendicular to the latter.

b) {\bf Oxygen bridge} (27 atoms): here the chain is interrupted by an Oxygen atom. Its spectrum (in Fig. 5) is dominated by  a feature at 7.5 $\mu$m, falling near the peak of the strongest UIB. In this mode, one of the two OC arms is stretching while the corresponding CH$_{2}$ group is rocking parallel to the C skeleton. The strength of this mode is partly due to the larger charge of the O atom. This structure will not be used as such, but is displayed here to illustrate the importance of oxygen bridges for our purposes. In the larger structures of Sec. 4, the O-bridge will appear in conjunction with much shorter carbon chains.

c) {\bf Chain CH$_{3}$a} (44 atoms) and {\bf Chain CH$_{3}$b} (35 atoms) : here, the hydrogen atoms of the methylene groups in type (a) are replaced by methyls. The two differ mainly by the shape of the carbon skeleton. The concatenation of their spectra (in Fig. 5) is dominated by a strong line at 8.5 $\mu$m, generated by a vibration that is essentially a rocking of H$_{3}$C-C-CH$_{3}$ groups parallel to the carbon skeleton. This concatenation will be taken to represent this topological class.

The dominant features of Chains CH$_{2}$ are much weaker than those of Chains CH$_{3}$; as a consequence, much less of the latter will be needed than of the former, in the sythetic spectra.

Unsaturated carbon chains will not be used as such, because they may enhance a stretching line at 5.2 $\mu$m, which is very weak in the sky. However, they may occasionally happen in some of the larger structures.

The chains naturally contribute to the aliphatic C-H stretching massif at 3.4 $\mu$m, whose strength relative to that of the aromatic C-H stretching band in the sky is highly variable: towards the Galactic Center, it even dwarfs the aromatic one. The abundance of chains is therefore an important tailoring factor.

\subsection{Aromatics}

\begin{figure}
\resizebox{\hsize}{!}{\includegraphics{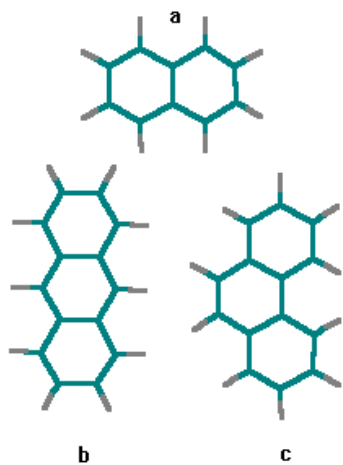}}
\caption[]{Small aromatics: naphtalene (18 at.), anthracene (24 at.) and phenantrene (24 at.).}
\end{figure}

Carbon chemistry also favours the combination of C atoms into rings, preferably hexagonal rings but also pentagonal, square and other rings. Naphtalene (18 at.), anthracene (24 at.) and phenantrene (24 at.; Fig. 2) are selected to represent part of this class. Their spectra are concatenated in Fig. 5. Their main contribution, via C-H out-of-plane bending vibrations, is to the 11-13 $\mu$m massif, and particularly to the strong 12.7 $\mu$m UIB.
Pyrene is not included because it has a strong line near 14.2 $\mu$m, which is not the case in celestial spectra. However, it will frequently appear within the larger structures of Sec. 4.

\subsection{Coronenes}
Coronene (36 at.) does not contribute much in the 5-10$\mu$m region, but it has very strong C-H stretching and bending oop (out-of-plane) features at 3.27 and 11.3 $\mu$m respectively. It also displays a strong feature near 17.9 $\mu$m, another UIB. Its high symmetry does not allow ir activation of more vibrations. 

\begin{figure}
\resizebox{\hsize}{!}{\includegraphics{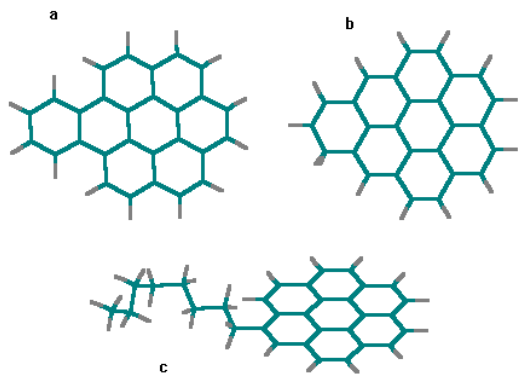}}
\caption[]{Coronene derivatives: a) one more hexagon is attached to one side of coronene proper, 42 at.; b) one more hexagon is attached to 2 sides of coronene proper, 41 at.; c) a tail (CH${2}$ chain) is added, 60 at.}
\end{figure}

However, its spectrum may be enriched by breaking this symmetry. Thus, attaching a new ring to one of the peripheral CC bond gives rise to a very strong feature at 12.8 $\mu$m, due to oop synchronous bending of the 4 CH bonds of this additional ring. Simultaneously, 8 strong, stretching lines are activated, spanning the range 3.25-3.32 $\mu$m. Similarly, a new cluster of lines are activated in the 6-7 $\mu$m range of C=C stretchings. Most interestingly, the 17.9 feature of coronene is replaced by 2 lines at 17.1 and 18.6 $\mu$m, also contributing to the 16-19 $\mu$m UIB massif. Again, the latter are due to bulk vibrational modes, with more or less symmetrical oop ring deformations.

Thus, slightly modifying the generic structure generates new types of features and/or contribute to the broadening of extant bands. Figure 3 displays 3 modifications (42, 41, and 60 atoms) of the coronene structure (36 atoms), which, together with the latter, generate the line spectrum displayed in Fig. 5, under the label ``coronenes". 

\subsection{Trios}

This label designates a 3-ringed type of structures often encountered in the analysis of kerogens: one pentagon between two hexagons; the free summit of the pentagon is often occupied by a heteroatom: oxygen, nitrogen or sulfur. 

\begin{figure}
\resizebox{\hsize}{!}{\includegraphics{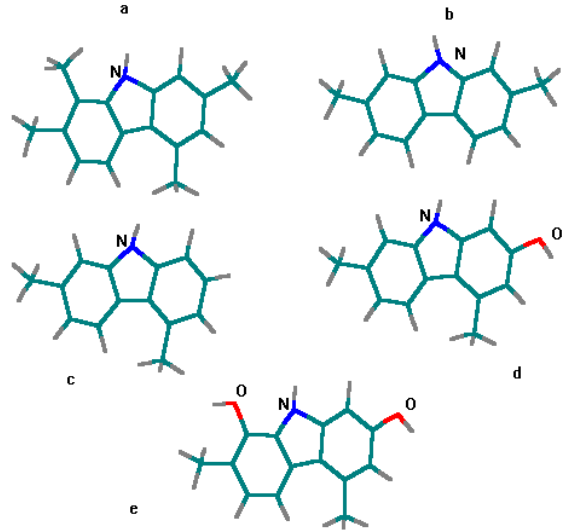}}
\caption[]{Nitrogen trios differing in details (namely, number and locations of CH$_{3}$ and OH terminations); 34, 28, 28, 29, 30 at from (a) to (e). }
\end{figure}

As this structure was found to be the most effective contributor to the UIBsbetween 15 and 20 $\mu$m, we considered several of its modifications, of which five were retained and displayed in Fig. 4. They comprise 34, 28, 28, 29 and 30 atoms, respectively. The concatenation of their spectra is also shown in Fig. 5 (upper graph).

Because of the higher electric charge of the heteroatoms, the latter generally yield more intense ir lines than C-capped trios. In particular, N-capped trios feature 3 groups of strong lines about 16, 18 and 19 $\mu$m. Their intensities decrease as N is successively substituted by O, S and C. The difference between N- and C-trios is illustrated in fig. 6. That is why only N-capped trios were retained in Fig. 4.

\bf OH group \rm : this group is occasionally attached to one of the structures. The reason for this is the relatively high cosmic abundance of oxygen and OH, and the high chemical activity of oxygen. Besides, several UIBs bear witness to the presence of oxygen: C=O stretching near 5.1 $\mu$m band, 
C-O and C=O stretchings, (C)OH bending in various configurations, between 6 and 9 $\mu$m (see below), and O-H wagging between 31 and 36 $\mu$m. Moreover, the presence of OH is found to considerably enhance ir activity in various regions of the spectrum, most spectacularly between 20 and 40 $\mu$m, where PPNe (ProtoPlanetary Nebulae) exhibit strong and wide bands (Kwok et al.\cite{kwo89}). Smith et al. \cite{smi07} also found evidence of enhancement of total UIB emission in metal-rich environments. Engelbracht et al. \cite{eng08}, analyzing the Spitzer Space Telescope data for 66 starbursting galaxies, found a similar trend, and even determined a threshold for the appearance of UIBs, at O/H$\sim10^{-4}$.

\begin{figure}
\resizebox{\hsize}{!}{\includegraphics{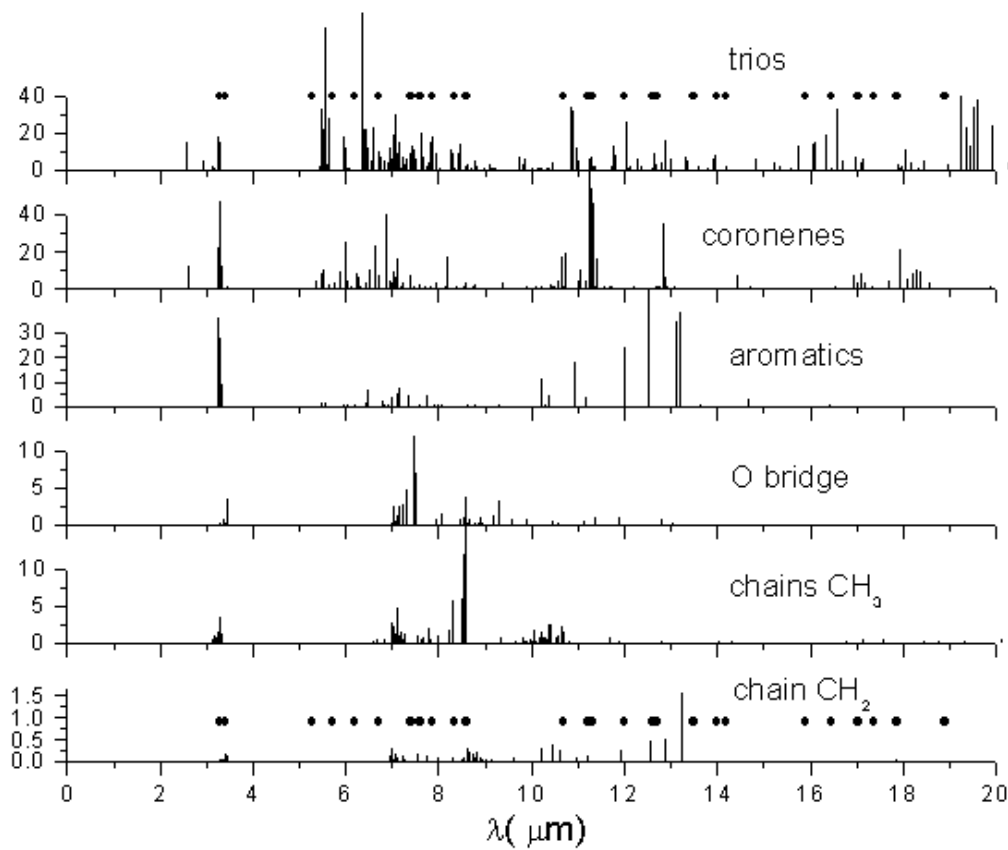}}
\caption[]{The single or concatenated spectra of ir intensities, $I$, linked to absorbances and bandwidths by eq. 1; from bottom up: Chain CH$_{2}$; Chain CH${3}$a and b; Chain with O-bridge; small aromatics; coronene and coronene derivatives a,b and c; trios a to e. The dots superimposed upon the top and bottom spectra each signal the peak of a UIB (Smith et al. \cite{smi07}). The spectral range is limited to 20 $\mu$m for clarity; contributions beyond that are negligible, except for trios (d) and (e), which carry hydroxyls, and therefore have lines near 34 $\mu$m (see text). }
\end{figure}

\begin{figure}
\resizebox{\hsize}{!}{\includegraphics{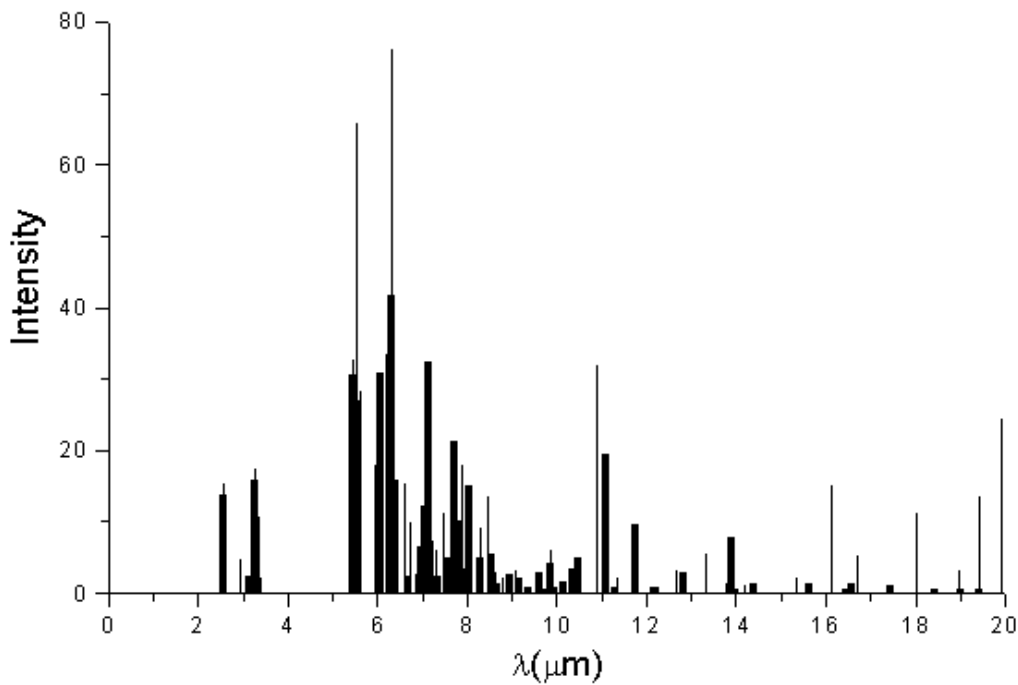}}
\caption[]{Spectral comparison between a N-capped trio (thin lines) and a C-capped trio (thick lines) derived from the former by substituting the N atom with a CH$_{2}$ group. Note the difference beyond 14 $\mu$m.}
\end{figure}

In Fig. 5 and 6, the spectral lines are represented by thin vertical lines whose lengths are proportional to their ir intensities. The intensity, $I$, of a mode is given in km.mol$^{-1}$, and 
should not be mistaken for the corresponding experimental and measurable quantity, the absorbance, which depends on the necessarily finite band width, itself being a function of the environmental parameters (e.g. temperature) and excitation process. The peak absorbance is given by

\begin{equation}
\alpha(\nu)=10^{2}\,\,I(\nu)\frac{C}{\Delta\nu},
\end{equation}

where $\alpha$ and $\nu$ are in cm$^{-1}$, $\Delta\nu$ is the bandwidth, and $C$ is the molar density in mol.l$^{-1}$. In fact, $I$ is a vector, parallel to the time derivative of the dipole moment; the software also delivers its 3 spatial components in the reference frame of the molecule.

\section{Notable ir bands}

Comparing the five spectra, one observes that each type of elementary structure contributes one or a few lines near the peak of one or the other UIB, indicated by a dot in the upper and lower spectra (from Smith et al.\cite{smi07}). Also, the lines displayed in Fig. 5 span the whole UIB range. The lines in each group can be assigned to similar vibrational modes, whose characteristics (stretching, bending, wagging, etc.) may be deduced from the observation of atomic motions, on the computer screen, as the structure is excited into a given mode. When several atoms (as opposed to small functional groups)  are set in motion in a given mode, the assignment is made easier by comparing the direction of motion of individual atoms with $\vec{I}$. This procedure highlights the role of heteroatoms in some transitions.

A few groups of lines may be distinguished for later comparison with the UIBs. Listed below are their \bf nominal \rm wavelengths, together with their range and assignment, whenever possible, to functional groups or bulk modes, for all computed structures (those retained here, as well as several others).

\bf 2.6 $\mu$m\rm

O-H stretching in the hydroxyl group; 2.57 to 2.60 $\mu$m.

\bf 2.9 $\mu$m\rm

N-H stretching; 2.91 to 2.92 $\mu$.

\bf 3.3 $\mu$m\rm

C-H stretching (aromatic); 3.25 to 3.33 $\mu$m.

\bf 3.4 $\mu$m\rm

C-H stretching (aliphatic); 3.38 to 3.44 $\mu$m.

\bf 5.1 $\mu$m\rm

C=O stretching; 5.07 $\mu$m.

\bf 5.2 $\mu$m\rm

C=C stretching in carbon chains; 5.17 to 5.19 $\mu$m. This is not displayed by saturated chains (as in Fig. 1), but in small chains bridging other elementary structures, as in some of the compacted trios of Sec. 4.1.

\bf 5.6 $\mu$m\rm

C-C and C=C stretching of the arms of rings in multiple-ring structures; 5.45 to 5.66 $\mu$m in trios, for which the intensities are strongest; 5.49 to 5.53 $\mu$m in coronene derivatives, where they are very weak, as noticed by Boersma et al. \cite{boe09}. Intensities are enhanced by the presence of hydroxyls. At the shortest wavelengths, two opposite sides of the central ring oscillate in phase; as the wavelength increases, they oscillate more and more out of phase.

\bf 6 to 8 $\mu$m\rm

In this range, a large number of modes are excited, involving essentially C-C and C=C stretchings which  display no obvious pattern: these are bulk modes whose frequencies depend on the size. Chains and common PAHs (naphtalene to coronene) contribute much less than trios. Intensities are considerably enhanced by the presence of attached hydroxyls.

A couple of stronger and narrower bands stand out in this range:

\bf 6.2 $\mu$m\rm

C-O stretching in COH groups, when the hydroxyl is attached to a 6-membered ring; 6 to 6.3 $\mu$m.
C-O stretching of one arm of an O-bridge between two aromatic groups is perturbed by the vibrations of the latter, and so can lie between 6 and 6.5 $\mu$m.
 
The 5.6- and 6.2-$\mu$m lines are very strong, particularly in isolated trios.

\bf 6.5 $\mu$m\rm 

C-N stretching in N-capped pentagons; 6.4 $\mu$m (asymmetric) to 6.6 $\mu$m (symmetric).
The C-O and C-N stretching lines are very strong; so, despite the low cosmic abundances of O and N relative to C, these atoms may contribute notably to the 6.2-$\mu$m UIB, as shown in Sec. 4. 

\bf 8.6 $\mu$m\rm

The only structures which display this band with enough strength are the aliphatic chain CH$_{3}$ (Fig 1c,d), even if no hydroxyl is attached; 8.24 to 8.59 $\mu$m. These modes are characterized by the motion of a carbon atom of the backbone, from perpendicular to the backbone at the shorterer wavelengths, to parallel at the longer ones. The intensities are relatively strong.

\bf 8 to 10 $\mu$m\rm

In this band, coronene and derivatives display weak lines associated with in-plane C-H bending vibrations.

\bf 10 to 13 $\mu$m\rm

As expected, this  band mostly displays strong, out-of-plane bendings of the C-H bonds of PAHs: as the wavelength increases, the motion involves more and more neighbouring bonds. In isolated benzene and pyrene, the vibration of all C-H bonds in phase gives rise to a strong ir line near 14 $\mu$m, which is not observed in the sky. These two species were therefore excluded from our selection.

The strongest lines in this band are due to solo C-H bending in coronene and its derivatives, and lie between 11.25 and 11.3 $\mu$m. These structures also display lines of various intensities between 12.6 and 12.9 $\mu$m. These two condensations of lines are reminiscent of the UIB peaks at 11.3 and 12.7 $\mu$m, while the lines due to the smaller aromatics, and distributed over the range, remind us of the so-called 11-13-$\mu$m plateau.

\bf 13 $\mu$m\rm

CH$_{2}$-chains have characteristic, weak modes between 12 and 13.3 $\mu$m, in which the methylene groups are rocking about an axis parallel to the chain's backbone. These must be taken into account in a model where such chains connect the other elementary structures.

\bf 15 to 20 $\mu$m\rm

The larger compact PAHs (coronene and derivatives) have bulk modes near 17 and 18 $\mu$m, in which the motions of the carbon atoms out of plane give rise to orderly ripples through the structure. The strongest band accurs in coronene at 17.9 $\mu$m, when only the H atoms and the sub-peripheral C atoms (those that are not linked to the H atoms) are set in motion, the two groups being out of phase. Trio structures have similar o.o.p modes near 17 and 18.1 $\mu$m; but they also contribute strong lines from 15.8 to 16.6 $\mu$m and near 19.5 $\mu$m. These three bands differ from one another by subtle changes in the relative phases of the motions of the C atoms. The motions and ir intensities are considerably enhanced when the central pentagon is capped with a N atom. Structure 4 a, for instance has a notable line at 15.8 $\mu$m, involving the motions, out of phase, of the N atom and its attached H. 

Moutou et al. \cite{mou99} previously suggested that 5-membered rings could be a component of the carrier of the IS 16.4 band they observed. Our finding that a N atom at the free apex of the pentagon enhances this group of lines is consistent with the observation of Smith et al. \cite{smi07} that the intensities of these lines increase with the metallicity of the environment.

Van Kerckhoven et al. \cite{van00}, Peeters et al. \cite{pee04} and Boersma et al. \cite{boe10} have shown that compact PAHs display several lines from 15 to 20 $\mu$m, which they also assign to C-C-C o.o.p bendings; but they feel that PAHs much larger than coronene are required for an acceptable fit to observations.

\bf 30 to 40 $\mu$m\rm

Wagging of the OH tail of the hydroxyl group attached to any other structure gives rise to strong lines in this range. The exact wavelength depends on the size of the structure and on the anchor point. For coronene, this occurs at 28.8 and 30.7 $\mu$m; for pyrene, at 36.4 $\mu$m.

When the hydroxyl is attached to one of the trios, its wagging feature occurs between 31 to 37 $\mu$m, and is usually stronger. Occasionally,  weak lines occur in this range even in the absence of OH, as in Fig. 4, type (a).

The assignments to vibrational modes, deduced from the observations of corresponding motions on the  screen, are seen to be in general agreement with organic chemistry tables.

This list includes most observed UIBs, as listed by Smith et al. \cite{smi07} for instance, which are represented by solid dots in the lower and upper graphs of Fig. 5. However, the ir-active lines are very sparsely distributed, even where they cluster together. The following reasoning helps quantifying this observation.

The line representation in Fig. 5 and 6 is independent on the
 broadening mechanisms. Now, a typical  width of the 3.3 $\mu$m UIB, for instance, is $\sim0.04\,\mu$m and even a spectral resolution of 1500 is not able to further resolve it (Tokunaga et al. \cite{tok}). If a Doppler broadening of 1/1000 dominates, then, at least 40 lines are required in order to fill only this band. H atoms are also involved in C-H in-plane and out-of-plane bending  and OH wagging modes. Several hundred H atoms are therefore needed to fill the other bands. If as many C atoms are included, we end up with at least a thousand atoms. 

Obviously, even all the selected structures taken together are not enough to produce a continuous spectrum. One way of increasing the number of lines within each UIB is to increase the size of elementary structures (length of chain, number of rings). But this becomes less and less effective as the size grows, as most new lines occur at longer and longer wavelengths (bulk modes). Also, this is hardly applicable to the trio structures. 

A more efficient way consists in slightly  modifying a given structure, by displacing a small group, e.g. methyl or hydroxyl, as was done for trios in Fig. 4. But the number of alternatives is rather limited for small structures.

On the other hand, evidence from a broad range of meteorites and organic matter in Comet 81P/Wild 2 samples does not point to abundant, large, PAHs; instead, it points to a synthesis of small reactive molecules, whose random condensation and subsequent rearrangement chemistry leads to a highly cross-linked macromolecule (see Cody et al. \cite{cod}). This is precisely the model upon which new structures were created below.

\section{Composite structures}
In order to produce composite structures, the elementary structures of Sec. 2 can be associated either by tightly connecting them with short and multiple bonds, or by concatenating them end to end with single bonds or small chains. A large number of structures were built along these lines.
Lack of space does not allow the inclusion of all illustrative sketches; only the leading member of each retained family is illustrated in the following figures. However, the following accompanying text is intended to stress that an increasing number of minor structural modifications of aptly selected composite structures, made up of the few molecules described in Sec. 2, and combined in the right proportions, will produce increasingly dense concentrations of lines in the right spectral bands.

\subsection{Compacted trios}
\begin{figure}
\resizebox{\hsize}{!}{\includegraphics{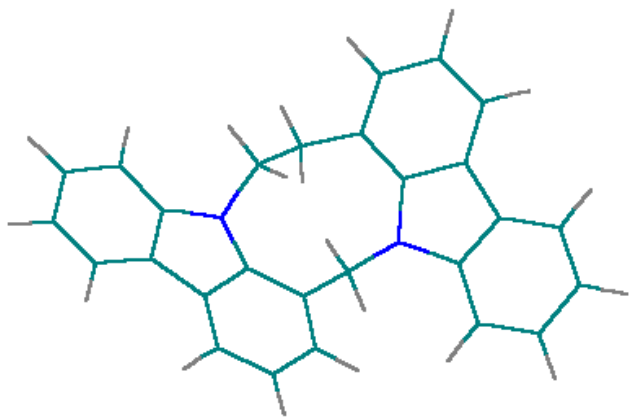}}
\caption[]{Compacted trios (a): 2 nitrogen trios (see fig. 4) tightly connected through their N atoms, 49 at.}
\end{figure}

Figure 7 shows a sample of this type: compacted trios (a), made of 2 nitrogen trios (see fig. 4) tightly connected through their nitrogen atoms (49 at.). The other samples to be used (but not shown here) are:

b) same as (a), except for the nitrogen atoms being substituted with C atoms and the attendant creation of two C=C bonds (47 at.);

c) structures (a) and (b) tightly connected (104 at.);

d) structure (c) plus 1 anthracene and 1 phenantrene (169 at.);

e) same as (d), except for another two short CH$_{2}$ links (190 at.).

The spectra of these 5 structures are concatenated into the lower spectrum of Fig. 11. As expected, the number of active ir lines notably increased as compared with isolated trios (upper spectrum, Fig. 5). The spectral density increased preferentially  within the width of a few UIBs. Besides, a budding ``grass" of weak but closely packed lines developed all over the spectral range beyond 6 $\mu$m, heralding the formation of a continuum beneath and between the UIBs. Each of these lines corresponds to a vibrational mode which involves neither a small and definite functional group, nor the whole structure. Such modes are intermediate between localized and bulk modes, in which many  atoms move in different directions, and therefore do not produce a large electric dipole moment; hence the weakness of these lines.

\subsection{Concatenated structures}

\bf Concatenated structures 1.\rm \, Member 1a of this family is shown in Fig. 8 (95 at.). This family, which contains no hydroxyls (OH), also includes 1b, which differs from 1a only by the displacement of a methyl group to another peripheral site.

\begin{figure}
\resizebox{\hsize}{!}{\includegraphics{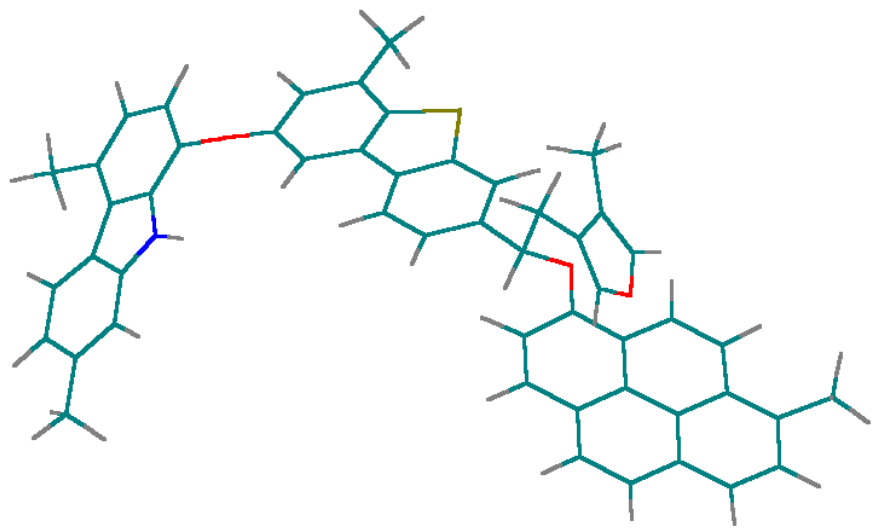}}
\caption[]{Concatenated structure 1a (95 at.).}
\end{figure}

\bf Concatenated structures 2.\rm  This family contains neither hydroxyls nor nitrogen. It is derived from sample 2a (93 at.; Fig. 9), which in turn is derived from concatenated structure 1a, by replacing N (in the N-capped trio) by CH$_{2}$. The family also includes

2b) 52 at.; derived from 2a by excluding the O-pentagon and the CH$_{2}$-capped trio, which leaves us with 1 pyrene and 1 S-capped trio, linked by a short chain -O-CH$_{2}$-;

2c) 49 at.; derived from (2b) by suppression of the CH$_{2}$ group in the short bridging chain;

2d) 43 at.; in (2c), substitute the pyrene with a Chain CH$_{2}$;

2e) 46 at.; in (2d), insert a CH$_{2}$ group between the O atom and the S-trio;

2f) 64 at.; in (2e), hang on a second CH$_{2}$ chain to the bridge between the S-trio and the first CH$_{2}$ chain.

2g)63 at.; in (2f), replace the second CH$_{2}$ chain by an O-capped pentagon.

\begin{figure}
\resizebox{\hsize}{!}{\includegraphics{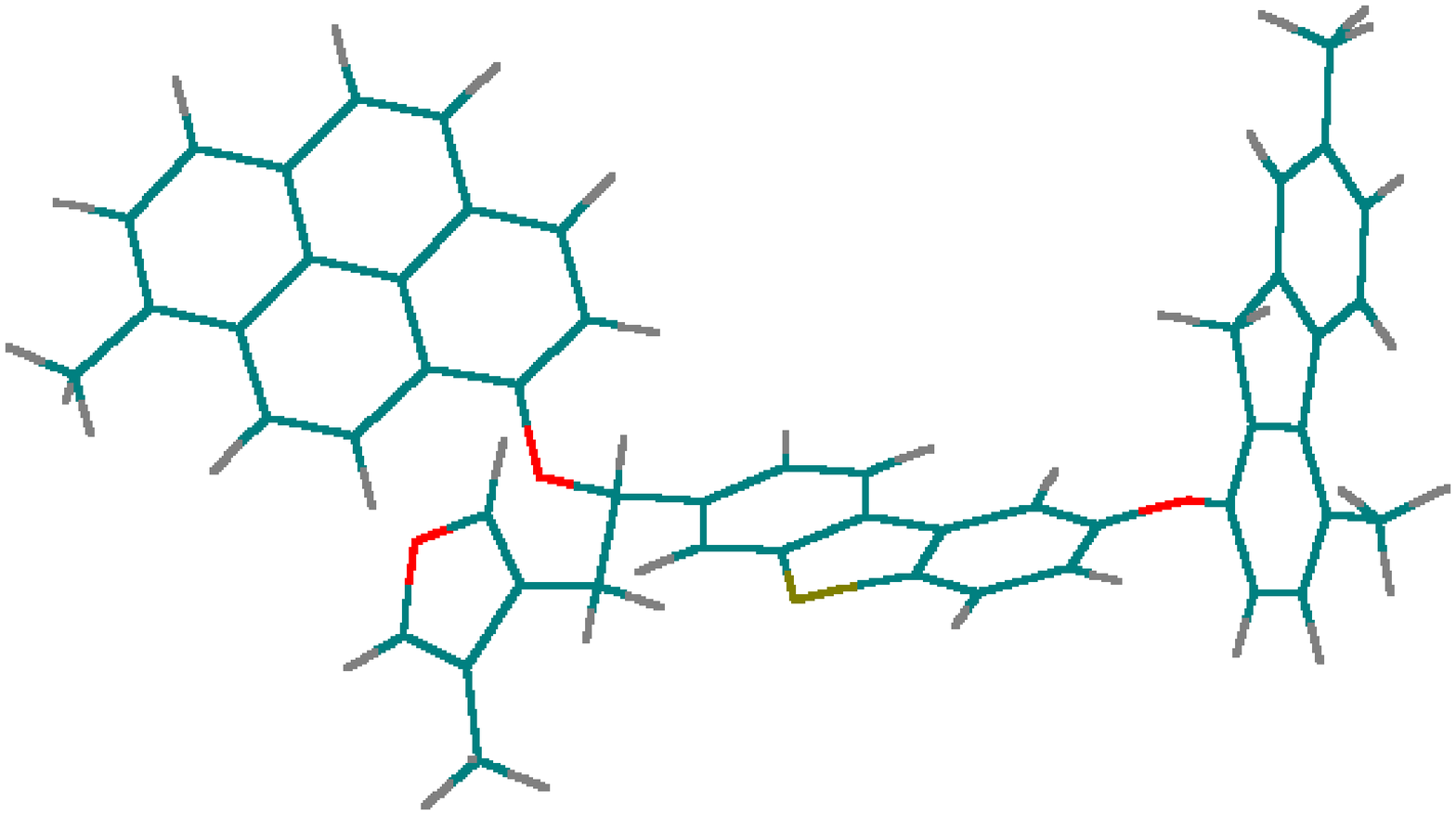}}
\caption[]{Concatenated structure 2a (93 at.).}
\end{figure}

\bf Concatenated structures 3.\rm  This family differs from the two previous one by the presence of several OH groups attached at the periphery of the main structure. The leading member (3a) is drawn in Fig. 10; 98 at.; other members are

3b) 97 at.; derived from (3a) by suppression of the OH group attached to the pyrene;

3c) 95 at.; derived from (3a) by substituting an H atom to the CH$_{3}$ group attached to the pyrene;

3d) 95 at.; in (3a), substitute an H atom to the CH$_{3}$ group attached to the O-capped pentagon;

3e) 95 at.; in (3a), substitute an H atom to the CH$_{3}$ group attached to the CH$_{2}$-capped trio;

3f) 102 at.; in (3a), replace the O-bridges with short CH$_{2}$ chains.

\begin{figure}
\resizebox{\hsize}{!}{\includegraphics{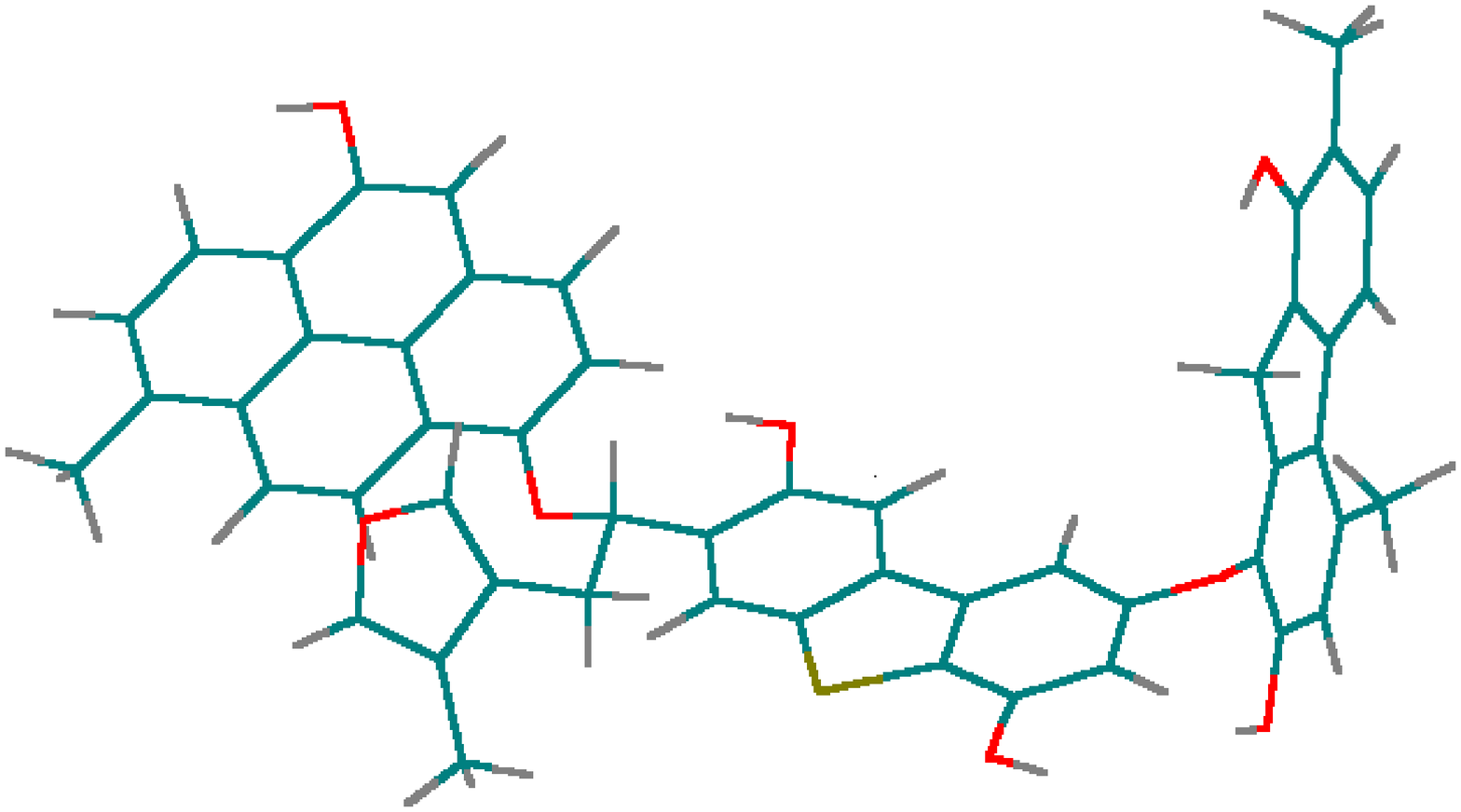}}
\caption[]{Concatenated structure 3a (93 at.).}
\end{figure}

Figure 11 displays the concatenated spectra of the 4 families of composite structures. Comparison with Fig. 5 demonstrates the efficiency of the adopted procedure in the way of increasing the density of active lines in the bands that are to be simulated. Although there is room for further improvement in this direction, e.g. by extending the list of families and variants in each family, it is instructive, at this stage, to try and produce a synthetic spectrum for comparison with observed UIBs.

\begin{figure}
\resizebox{\hsize}{!}{\includegraphics{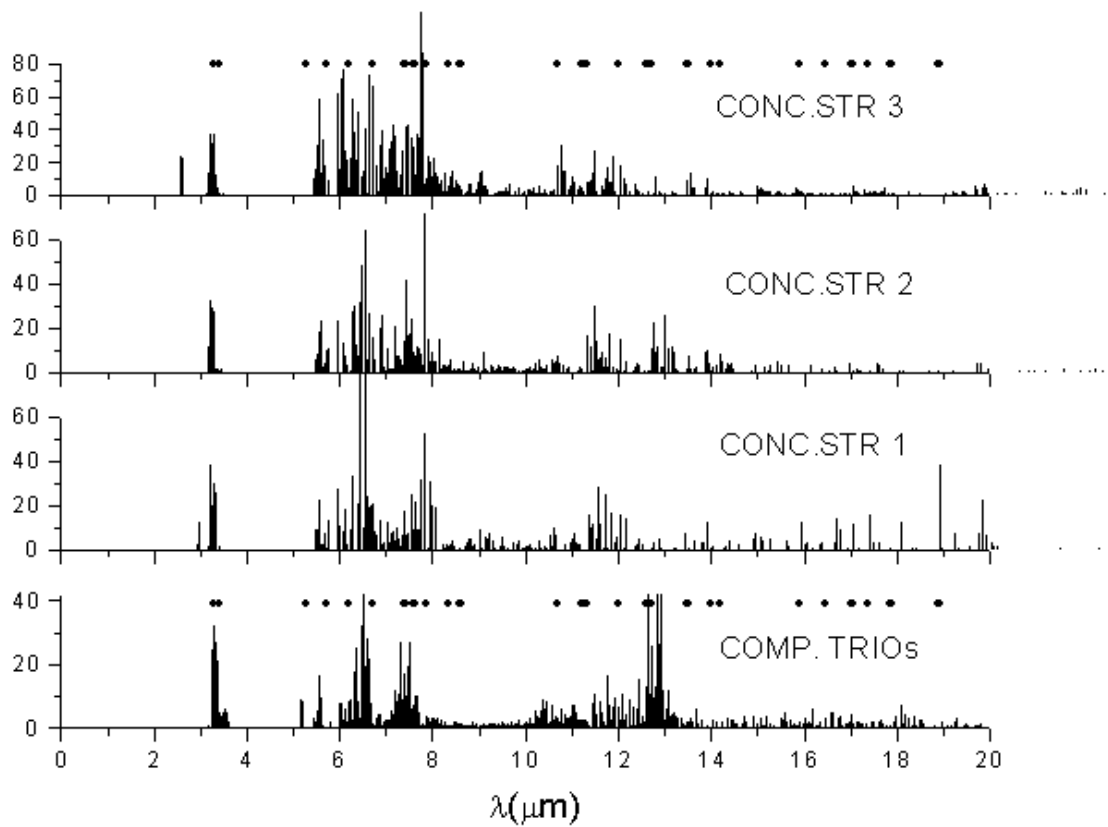}}
\caption[]{The concatenated spectra of ir intensities, $I$; from bottom up: compacted trios (a) to  (e); concatenated structures 1a and 1b; concatenated structures 2a to 2g, concatenated structures 3a to 3f. The spectral range is limited to 20 $\mu$m for clarity; contributions beyond that are negligible, except for concatenated structures type 3, which carry hydroxyls, and therefore have lines from 30 to 40 $\mu$m (see text).}
\end{figure}

\section{Model synthetic spectra}

In the dust model envisioned here, the UIB spectrum is the sum of contributions from all the structures described above. Although, the emission spectrum depends on the excitation process, we assume, here, for simplicity, that it is predominantly determined by the intensity $I$, which is proportional to the absorbance of the structure. Then, the contribution of a given structure, at  a given wavelength, is the product of the corresponding line intensity, $I$, in its spectrum, with a multiplying factor, $f$, representing its relative abundance in the dust. The same factor applies to all members of a given family of structures, and the concatenation of these products constitute the model spectrum. Taking as a benchmark the spectrum of NGC 1482 displayed in Smith et al. \cite{smi07}, and the nominal feature wavelengths listed by the same authors, we arrived, by trial and error, at the $f$ factors in the last column of Table 1.

\begin{figure}
\resizebox{\hsize}{!}{\includegraphics{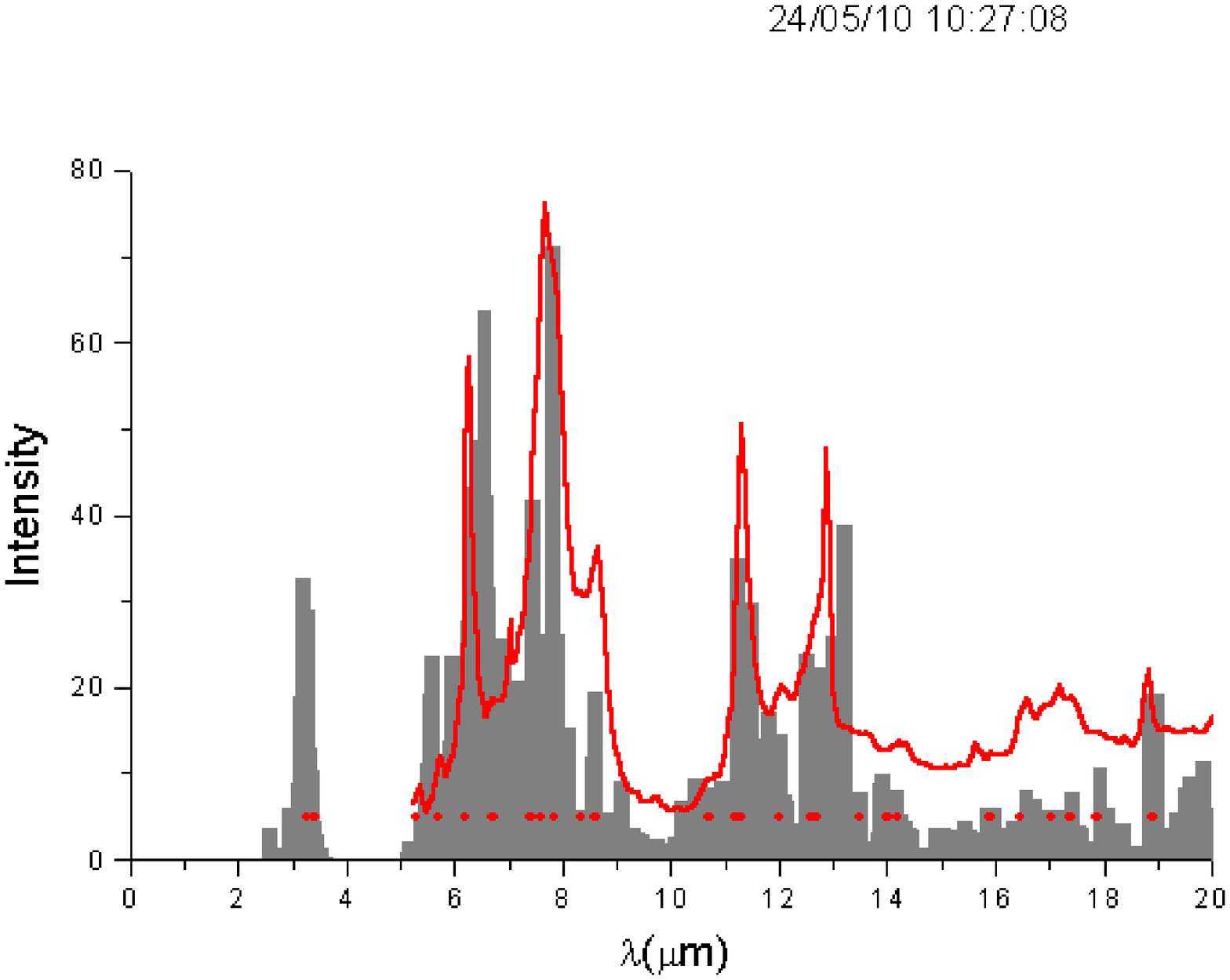}}
\caption[]{A model synthetic spectrum for star-forming galaxies: this is a concatenation of the spectra of the various structures in Sec. 2 and 4, weighted by the $f$ factors of Table 1. The width of each individual line is arbitrarily set at 0.2 $\mu$m for clarity. The red curve is the spectrum of NGC 1482, the red dots stand each for one of the observed peaks, all reproduced from Smith et al.\cite{smi07}; dots were added at 3.27 and 3.4 $\mu$m, for the CH stretching UIBs.}
\end{figure}

\begin{figure}
\resizebox{\hsize}{!}{\includegraphics{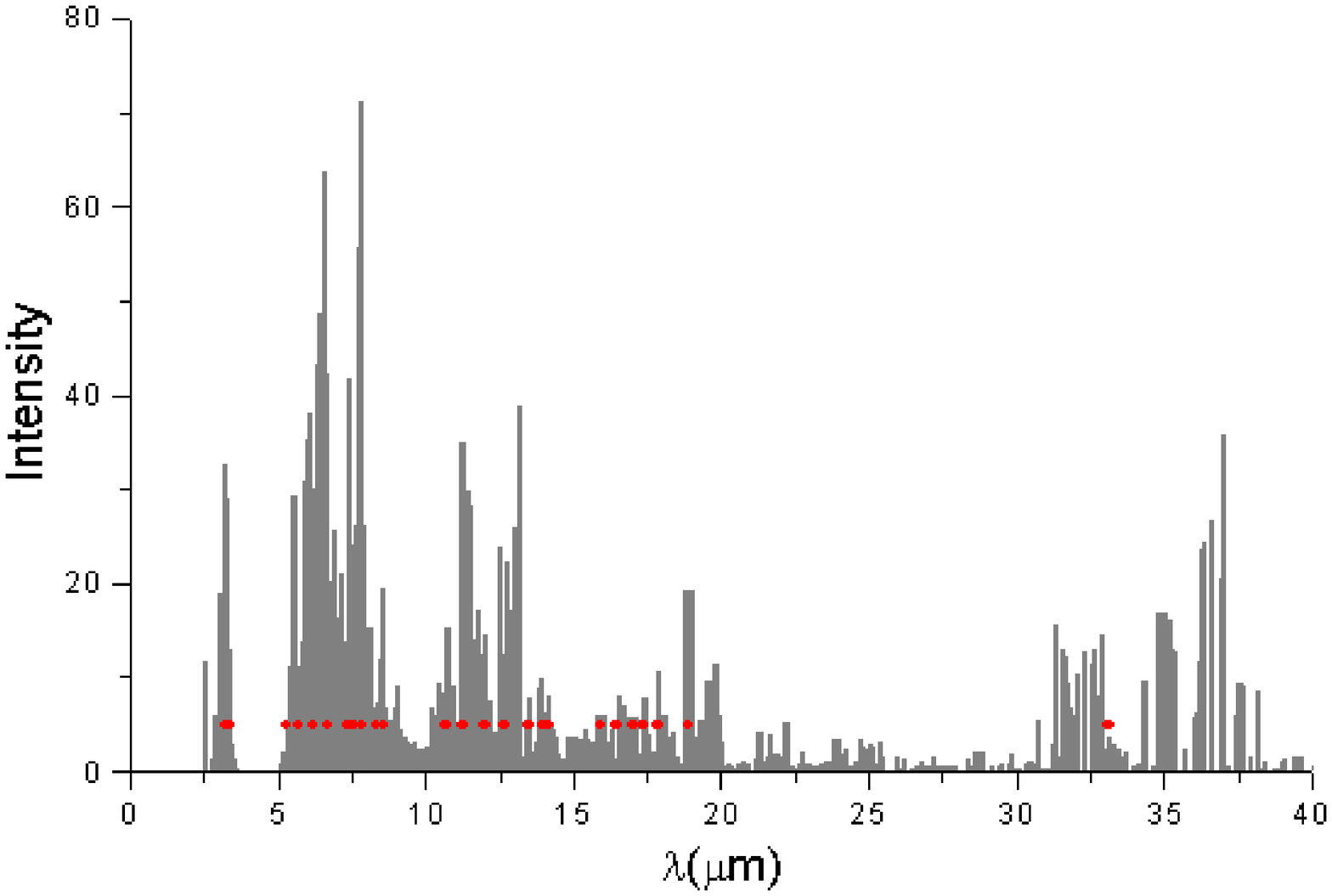}}
\caption[]{A model synthetic spectrum for PPNe: this is a concatenation of the spectra of the various structures in Sec. 2 and 4, weighted by the same $f$ factors as in Fig.12, except for concatenated structures 3, with $f=1$ here. The main effect is the emergence of a wide and strong band between 30 and 40 $\mu$m, to be compared with the FIR band described by Kwok et al. \cite{kwo89}. The red curve is the spectrum of NGC 1482, the red dots stand each for one of the peaks, all reproduced from Smith et al. \cite{smi07}; dots were added at 3.27 and 3.4 $\mu$m, for the CH stretching UIBs.}
\end{figure}

The concatenated spectrum obtained in this way is presented in fig. 12, for the case where $f$(concatenated structures 3)=0. For the sake of clarity, the maximum wavelength was limited at 20 $\mu$m and the line width was uniformly set at $\sim0.2 \mu$m. Most lines then overlap their neighbours. It must be stressed that the software used here was optimised for small hydrocarbon molecules and, therefore, does not pretend to a wavelength accuracy better than a few percent, especially when applied to large structures, and when heavy heteroatoms are included. 

Notwithstanding these reservations, Fig. 12 does exhibit the general profile of the celestial spectra: the stronger peaks do not deviate from the observed positions by more than 6$\%$; nearly all observed peaks are present with approximately the right relative intensities; the continuum is distinctly building up as the number of structures in each family increases, and no strong undesirable feature emerges. 

Some defects are, however, obvious. The lines at 5.6 and 19 $\mu$m, both associated with trios,  appear to be too strong; this suggests that trios should not carry hydroxyl groups, which enhance those features. The peak at 6.55 $\mu$m is too red by 6 $\%$; this, unfortunately cannot be remedied with the present software. The line density in the 7.7-$\mu$m complex is insufficient, as is also the case near 12.5 $\mu$m, indicating that more, and different, structures are required, so as to provide more modes, etc. 

Not shown in Fig.12 are 2 isolated lines near 34.3 $\mu$m ($I$=10), due to OH wagging in trios (d) and (e). They can be seen in Fig. 13. The spectrum in the latter figure was obtained by adding to the concatenation of Fig. 12 the spectra of concatenated structures (3a) to (3f), which include many more hydroxyls, attached to a greater variety of substructures. As a result, the wagging frequencies now cover a much broader band, which probably corresponds to the 33.1-$\mu$m listed by Smith et al. \cite{smi07}, and to the broad FIR bands documented by Kwok and coll. (see Hrivnak et al. \cite{hri09}, and bibliography therein).

\section{Model dust composition and structure}

\begin{table*}
\caption[]{Atomic composition of dust}
\begin{flushleft}
\begin{tabular}{lllllllll}
\hline
Name & N$_{at}$ & C & H & O & N & S & OH & f\\
\hline
Chains CH$_{2}$ & 29 & 9 & 20 & 0 & 0 & 0 & 0 & 25\\
\hline
Chains CH$_{3}$ & 79 & 25 & 54 & 1 & 0 & 0 & 0 & 1\\ 
\hline
Coronenes & 179 & 111 & 68 & 1 & 0 & 0 & 0 & 0.5\\
\hline
Aromatics & 66 & 38 & 28 & 0 & 0 & 0 & 0 & 0.5\\
\hline
Trios & 149 & 72 & 69 & 3 & 5 & 0 & 3 & 0.25\\ 
\hline
Compacted trios & 559 & 315 & 236 & 0 & 8 & 0 & 0 & 0.25\\
\hline
Conc. struct. 1 & 190 & 102 & 78 & 6 & 2 & 2 & 0 & 0.5\\
\hline
Conc. Struct. 2 & 400 & 199 & 184 & 10 & 0 & 6 & 0 & 1\\
\hline
Conc. struct. 3 & 572 & 305 & 226 & 45 & 0 & 6 & 29 & 0/1\\
\hline
Total & 2183 & 977 & 1127 & 63 & 3 & 13 & 30 & 1\\
\hline
\end{tabular}
\end{flushleft}
\end{table*}

Table 1 lists the atomic composition of the various families of structures included in the model dust which deliver the spectra of Fig. 12 and 13. The last column gives the corresponding abundance fractions, $f$. The last row lists the total numbers of atoms in the case of Fig. 13, where the concatenated structures type 3 are included. The relative number abundances are:

$H/C=1.15,\,\,O/C=6.4\,10^{-2},\,\,N/C=2.6\,10^{-3},\,\,S/C=1.3\,10^{-2}$.

The fraction of C atoms in the whole population is 0.45. Of all the O atoms, the fraction in the form of hydroxyl is 0.48; the rest is in bridges and 5-membered rings. Interestingly, the hydrogen and oxygen ratios are characteristic of kerogens type II in advanced stages of dehydrogenation (see the Van Krevelen diagram reproduced in Papoular \cite{pap01}), but still poor graphitization. 
This highlights the prospects of simulating UIB spectra in the laboratory, using easily available and well characterized kerogens.

While NGC 1482 is typical of star-forming galaxies, large variations of the relative feature intensities are observed from galaxy to galaxy (see Smith et al. \cite{smi07}), and all the more so for other types of galaxies, for PNe, PPNe (see Kwok et al. \cite{kwo89}), reflection nebulae and even from site to site within these objects (see Sellgren et al. \cite{sel07}; Werner et al. \cite{wer04}). The differences between the spectra of our selected structures (Fig. 5 and 11; Fig. 12 and 13) suggest that tailoring the factor $f$ can reproduce at least part of the variety of astronomical spectra. 

It was implicitely assumed, above, that the model dust consists of sets of individual structures, chosen among the 35 described in Sec. 2 and 4, with variable relative numbers tailored so as to fit  different observed spectra, and taking into account the differences in the spectral contributions of different structures, as illustrated in Fig. 5 and 11. However, there is also the possibility that, depending on their evolutionary stage, some of these structures coalesce into larger grains. Figure 7 is an example of tight binding between two trios. Comparison of the upper spectrum of Fig. 4 and the lower spectrum of Fig. 7 shows that the change may be considerable. Not so for loose coupling, an example of which is a ring of structures chosen among those selected in Sec. 2 and 4, and linked to one another by chains of at least 10 CH$_{2}$ groups, with no more than 2 chains connected to each structure. Chemical computer experiments show that, by comparison with the concatenation of the spectra of the individual structures included in the ring, the following changes occur:

a) more lines become ir active, thus increasing the spectral line density;

b) lines of the constitutive structures are slightly shifted ($<$ a few $\%$);

c) new modes and lines are created, mostly weak and farther in the ir.

If only a chain of 10 CH$_{2}$ groups is attached to one individual structure, the spectral change is generally limited to line shifts of less than 0.1 $\mu$m. These changes are understandable in terms of perturbation by coupling, decreased perturbation as the chain lengths increase, weak ir contribution of the chains as such (see lower spectrum of Fig. 5), and creation of bulk modes, whose wavelengths increase with the size of the composite structure. Overall, coupling structures into grains enhances ir activity for a given quantity of matter.

Pursuing in this vein, and again taking our clue from kerogens, we assume the density of carrier material, built ``loosely" in this way, to be of order 1 g.cm$^{-3}$. The developments above suggest that the minimum number of C atoms required in a grain of this material for it just to begin mimicking a UIB spectrum, is about 1000 and certainly more like 10$^{4}$, for the spectrum to be continuous at the level of 0.1 $\mu$m. The minimum diameter of such a grain is, therefore, of order 100 $\AA{\ }$. 

It must be remembered that the celestial spectra generally include an underlying continuum which rises towards the FIR. Our model is not designed to reproduce this continuum, although the jamming of weak lines within the UIBs, and the bulk modes referred to above, in large structures, may contribute to it. Apart from minerals, like silicates, which are ubiquitous, graphitic particles, which are responsible for the 2175 \AA{\ } band, are also important contributors. Such graphitic matter is a natural outcome of the carbonization and graphitization processes (amply documented in the earth and in the laboratory) which, in time and under radiation and shocks, liberate the volatiles from the kerogen-like structures of our model, then closes chains into rings, and finally concatenate aromatic rings into densely packed graphitic flakes. While such grains may have an independent existence, it may also be envisioned that a typical dust grain may consist of stacked onion layers corresponding to different stages of this evolution, with the most evolved at the center, and the less evolved being continually deposited at the periphery.

\section{The excitation process}

\begin{figure}
\resizebox{\hsize}{!}{\includegraphics{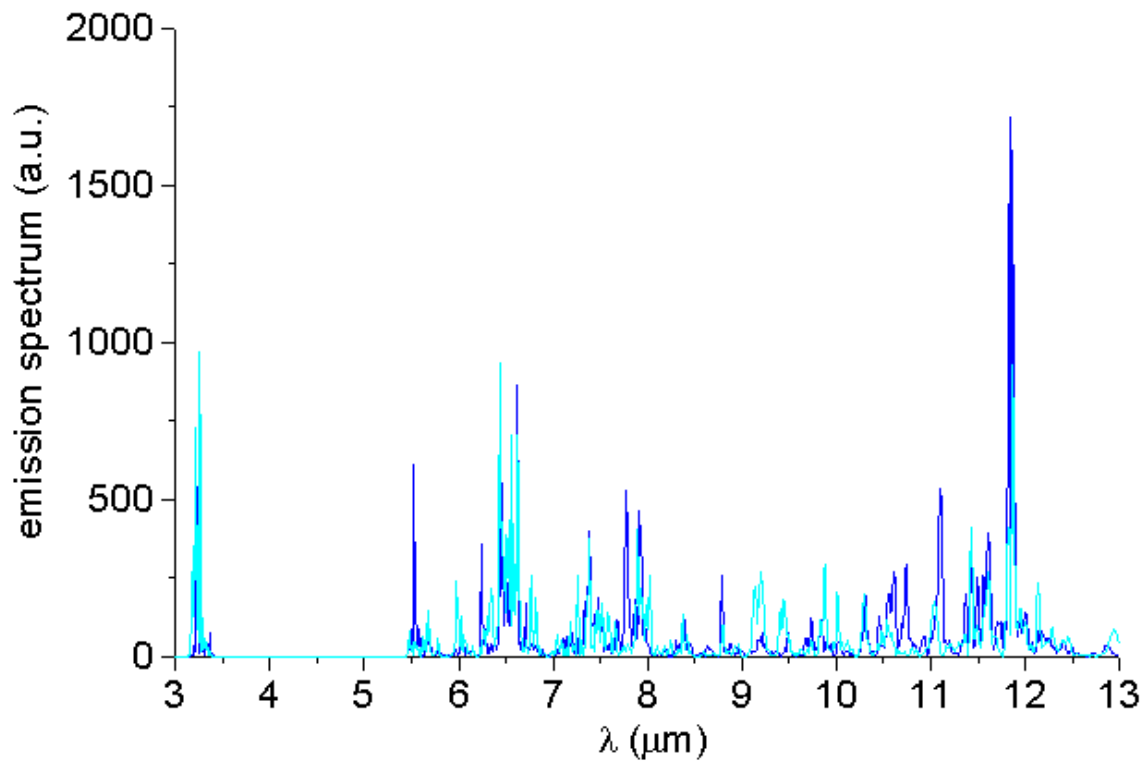}}
\caption[]{a)(blue): the emission spectrum (a.u., see text) of the structure drawn in Fig. 8, deduced from the computed molecular dynamics during 10 ps, starting 23 ps after an impinging H atom was captured by a dangling bond at the periphery of the particle. This represents the average, over 10 ps, of the probability of emission of photons by the various vibration modes after relaxation of the local perturbation; b)(cyan): the same, after H capture at another site. Line broadening due to finite computer run time: $\sim0.03 \mu$m.}
\end{figure}

In principle, the spectra of ir activity (or intensity, $I$) given above are necessary, but not sufficient to define unambiguously ir absorption or emission by a given structure. For absorption, one also needs to know the bandwidth (eq. 1), which depends essentially on the ambient temperature, through the Doppler effect. However, in the interstellar medium, the temperature is generally so low that the broadening is negligible as shown at the end of Sec. 3; the thin vertical lines in the ir activity spectra are then a good approximation to the absorption spectra, to a multiplicative factor.

As for emission, it also depends on the particular excitation process in action. It is generally admitted that the UIB emitters, in the interstellar medium, cannot be heated to thermal equilibrium temperatures high enough that their shortest wavelength bands be adequately excited. On the other hand, in the present dust model, as in most others, an acceptable spectral fit to observations requires grain sizes in the order of 100 atoms or more. I have already argued in detail why such grains are not likely to be adequately excited by the available single UV photons, as in the stochastic heating model (Papoular \cite{pap05}). I proposed chemiluminescence as an alternative (op. cit.). In this model, UV radiation only plays an indirect role, in producing a ubiquitous population of H radicals which are the direct exciting agent. Indeed, they constantly interact with the peripheral hydrogens of the kerogen-like grains, either abstracting one H atom to form a hydrogen molecule, and leave behind a dangling C bond, or by being captured by one such bond  (Papoular \cite{pap05b}). In both cases, the grain atoms are set in motion in an attempt to readjust the structure, and, in the very thin interstellar medium, collisions are extremely rare so the corresponding energy can only be evacuated by ir radiation. It is found that H capture by dangling bonds (chemisorption) leaves much more energy in the grain than does abstraction, so the latter is not considered further.

The energy gained by the grain in chemisorption is the C-H bond energy ($\sim$5 eV). It is initially localized in the newly created bond in the form of a stretching vibration mode. Its fate afterwards has been studied numerically and experimentally in several papers (see Papoular \cite{pap06}), and bibliography therein). It turns out that energy is preferentially funnelled to those other modes that are more strongly coupled to the C-H stretching modes, and so forth down the coupling ladder. When dynamical equilibrium finally sets in, after 0.01 to 1 ns (depending on the particle size), all modes are excited, but the energy partition between them is neither uniform nor thermal. Roughly speaking, the  ir-active modes are not excited uniformly, and the excited modes are not uniformly ir-active. At a given time, the probability for a vibration mode to emit an ir photon is proportional to the energy stored in it, but also to its intensity $I$ (or ir activity, eq. 1). It also depends on the particular site of H capture. However, the energy distribution among modes keeps changing in time because of exchanges between coupled modes, so that all accessible modes may eventually be visited before the photon is emitted (up to 1 sec radiation time).

As an illustration, the concatenated structure in Fig. 8 was perturbed by stretching one of its CH bonds from its length in the ground state, 1.096 \AA{\ }, to 2.2 \AA{\ }, thus simulating H capture by a dangling bond. This imparts to it an excitation energy 124 kcal.mol$^{-1}$, as is the case for H capture by a dangling bond. It was then left to relax and settle in the dynamical equilibrium state. This is the case when the variance of the total kinetic energy (for instance) ceases to decrease and stays constant on average. This state was reached after $\sim$10 ps. The run was stretched out to 33 ps and the last 10 ps retained for spectral analysis. The resulting energy distribution among the modes was obtained by considering the kinetic energy of the system, extracting its oscillating part, computing the auto-correlation function of the latter, and taking the FFT of this function. The relative probability of photon emission by the various vibrations was obtained by multiplying this energy spectrum by the spectrum of line intensities as computed in Sec. 4. This result is shown in blue in Fig. 14. Superimposed upon it (in cyan) is a spectrum obtained in the same way, except for a different site of H capture. Note that the peaks coincide in wavelength, as expected, but differ in height. Comparison of this graph with the spectrum of line intensities, $I$, of the structure under consideration (see Fig. 11) shows that the intensity spectrum is an operational approximation to the emission spectrum too.

Also note the narrowness of the peaks, which is limited by the small time length of the data sample, not by the equivalent temperature, which oscillates around $\sim250$ K. Clearly, the synthesis of a model spectrum approaching observed spectra requires a great many different structures and a large number of H captures at different sites, which justifies our quests for structures in Sec. 4 and 5.

The dynamics of a molecule maintained in a thermal bath at a resonable temperature was also computed with the same software. It was found that the energy distribution among the normal modes was neither thermal nor even regular, let alone uniform. The emission spectrum has the same general characters as those of Fig. 14. The sparsity of spectral peaks delivered by both excitation models may explain the relatively small number of UIBs observed between 2 and 20 $\mu$m.

\section{Conclusion}

Using only a few small, carbon rich, molecular building blocks (aliphatic chains, aromatics, pentagons sandwiched between hexagons, attached methyl and hydroxyl functional groups), we have obtained a variety of ir spectra, each displaying one or more lines within the observed UIBs. Clusters of ir lines, at the right wavelengths, are created by slightly modifying the structural details, e.g. the number and location of functional groups, and by associating the elementary structures through aliphatic chains to form loosely packed grains. Combining these spectra together in adequate proportions can deliver different spectra, all exhibiting the UIBs, but with different relative intensities, so as to simulate emissions from different environments.

 A very large number of spectra are needed so that, in their concatenated spectrum, the spectral density within each cluster of lines be so high as to mimic a continuous band. Here, a total of about 6000 lines were concatenated and more can be obtained by simple extensions of each of the 8 considered structural families, or by introducing larger/different elementary structures.

Interestingly, all our spectra were obtained with neutral molecules and no dangling bonds. Also, it was possible to assign different, characteristic, vibrational modes to each of the model UIBs. This may help in designing new, relevant, structures. 

The probability for a vibration mode to emit an ir photon is proportional to its intensity $I$ (or ir activity, eq. 1), but also  to the energy stored in it. The latter depends on the excitation process.  As an example, the case of excitation by H atom capture was studied in some detail. The resulting emission spectrum does not differ essentially from the intensity spectrum.

\section{Acknowledgments}
I am deeply grateful to Prof. Sun Kwok for sharing data and information, and for his sustained support. I am also indebted to Dr J. D. Smith for kindly providing useful data files.

\end{document}